\documentclass[aip,apl,amsmath,amssymb,preprint,]{revtex4-1}
\usepackage{here}
\usepackage{graphicx}
\usepackage{bm}
\usepackage{times}
\usepackage{xcolor}
\usepackage{multirow}
\usepackage{comment}
\usepackage{braket}

\usepackage{dcolumn}% Align table columns on decimal point

\usepackage{ulem}%

\usepackage[utf8]{inputenc}
\usepackage[T1]{fontenc}
\usepackage{mathptmx}
\usepackage{etoolbox}

\makeatletter
\def\@email#1#2{%
 \endgroup
 \patchcmd{\titleblock@produce}
  {\frontmatter@RRAPformat}
  {\frontmatter@RRAPformat{\produce@RRAP{*#1\href{mailto:#2}{#2}}}\frontmatter@RRAPformat}
  {}{}
}%
\makeatother

\begin{document}

\preprint{AIP/123-QED}

%-------------------------------------------------------------------------------------------------------          Title          ---------------------------------------------------------------------------------------------------------------------
\title
{Fe contribution to the magnetic anisotropy of $L{\rm1_0}$-ordered FePt thin films studied by angle-dependent x-ray magnetic circular dichroism}

\author{Goro Shibata}
\affiliation{Department of Physics, University of Tokyo, Tokyo 113-0033, Japan}
\affiliation{Materials Sciences Research Center, Japan Atomic Energy Agency, Sayo 679-5148, Japan}
\email{shibata.goro@jaea.go.jp}

\author{Keisuke Ikeda}
\affiliation{Department of Physics, University of Tokyo, Tokyo 113-0033, Japan}

\author{Takeshi Seki}
\affiliation{Institute for Materials Research, Tohoku University, Sendai 980-8577, Japan}

\author{Shoya Sakamoto}
\affiliation{Department of Physics, University of Tokyo, Tokyo 113-0033, Japan}
\affiliation{Institute for Materials Research, Tohoku University, Sendai 980-8577, Japan}

\author{Yosuke Nonaka}
\affiliation{Department of Physics, University of Tokyo, Tokyo 113-0033, Japan}

\author{Zhendong Chi}
\affiliation{Department of Physics, University of Tokyo, Tokyo 113-0033, Japan}

\author{Yuxuan Wan}
\affiliation{Department of Physics, University of Tokyo, Tokyo 113-0033, Japan}

\author{Masahiro Suzuki}
\affiliation{Department of Physics, University of Tokyo, Tokyo 113-0033, Japan}

\author{Tsuneharu Koide}
\affiliation{Institute of Materials Structure Science, High Energy Accelerator Research Organization (KEK), Tsukuba 305-0801, Japan}

\author{Hiroki Wadati}
\affiliation{The Institute for Solid State Physics, The University of Tokyo, Kashiwa 277-8581, Japan}
\affiliation{Graduate School of Material Science, University of Hyogo, Sayo 678-1297, Japan}

\author{Koki Takanashi}
\affiliation{Institute for Materials Research, Tohoku University, Sendai 980-8577, Japan}
\affiliation{Advanced Science Research Center, Japan Atomic Energy Agency, Tokai 319-1195, Japan}

\author{Atsushi Fujimori}
\affiliation{Department of Physics, University of Tokyo, Tokyo 113-0033, Japan}
\affiliation{Center for Quantum Science and Technology and Department of Physics, National Tsing Hua University, Hsinchu 300044, Taiwan}

\date{\today}

\begin{abstract}
Among magnetic thin films with perpendicular magnetic anisotropy (PMA), $L1_0$-ordered FePt has attracted significant attention because of its exceptionally strong PMA. However, the microscopic origin of its strong PMA has not been elucidated experimentally.  We have investigated the contribution of the Fe $3d$ electrons to its magnetic anisotropy energy by angle-dependent x-ray magnetic circular dichroism at the Fe $L_{2,3}$ edge. 
By this technique, one can deduce the magnetic dipole moment $m_\text{T}$, which represents the anisotropic spatial distribution of spin-polarized electrons, and the orbital moment anisotropy (OMA) of Fe $3d$ electrons. 
Detected finite $m_\text{T}$ indicates that the spin-polarized Fe $3d$ electrons are distributed preferentially in the out-of-plane direction of the films. 
This $m_\text{T}$ of Fe overwhelms the positive contribution of OMA to PMA, and reduces the PMA of $L1_0$-ordered FePt thin films, 
consistent with a previous first-principles calculation. 
The present result implies that a large positive contribution of the non-magnetic element Pt rather than Fe governs the PMA of $L1_0$-ordered FePt thin films. 
\end{abstract}

\maketitle

%\graphicspath{{fig/}}

%%%%%%%%%%%%%%%%%%%% INTRODUCTION

Magnetic thin films with perpendicular magnetic anisotropy (PMA) have been extensively studied since they are essential for applications in high-density magnetic recording media and novel spintronics devices. \cite{PMAreview_RMP2017, PMAreview_Vacuum2017, Miwa_VCMAreview_JPhysD2019, VCMAreview_JMMM2022} 
Among them, $L1_0$-ordered FePt has drawn particularly strong attention because it shows PMA with exceptionally high magnetocrystalline anisotropy (MCA) energy %$K_\text{u} 
of $\sim 10^6$--$10^7$ J/m$^3$. \cite{HDD_FePt,FePt_Seki_fund,OrderedAlloy_Seki} 
By adjusting the deposition and post-annealing temperatures, one can control the degree of the long range chemical order ($S$) in a wide range, \cite{FePt_Seki_fund, Shima_APL2002} a useful property for realizing desired device characteristics. 
Voltage-controlled magnetic anisotropy %(VCMA) 
of the $L1_0$-ordered FePt thin films 
has also been a topical issue in recent years. \cite{Weisheit_VCMA_FePt_Science2007, Seki_voltage_APL2011, Miwa_VCMAreview_JPhysD2019, Nozaki2019_VCMA_review, VCMAreview_JMMM2022, Miwa_2017aa} 

Despite the concentrated effort over decades, the origin of the huge PMA of $L1_0$-FePt has not been fully elucidated. 
In general, the MCA of ferromagnetic thin films is expressed as a sum of two components within the second-order perturbation theory with respect to the spin-orbit coupling (SOC): \cite{Brunoeq, Wang_spinflip, vanderLaan, SuzukiMiwa2019} 
One is the term proportional to the anisotropy of the orbital magnetic moment ($m_\text{orb}$) between the out-of-plane and in-plane directions of the film, referred to as orbital moment anisotropy (OMA), \cite{Brunoeq} and the other is the term which reflects the asphericity of electron orbitals (i.e.\ whether the orbitals are elongated or shrunk along the out-of-plane axis) induced by SOC. \cite{Wang_spinflip, vanderLaan, SuzukiMiwa2019} 
In the case of $L1_0$-FePt, there have been controversies on which of these two terms contribute more strongly to the MCA and which of Pt and Fe has dominant contributions. 
Experimentally,  the OMAs of both Fe \cite{Soares_FePtOMA,FePt_Ikeda} and Pt \cite{FePt_Ikeda,Miwa_2017aa} have been revealed by x-ray magnetic circular dichroism (XMCD). 
However, some theoretical calculations show that the MCA energy from Fe is negligibly small or even negative despite the large OMA. \cite{Solovyev_RSGreens_FePt,Ueda_spinflip_HXPES_FePt} 
In addition, although many studies suggest that the Pt contribution is dominant in the MCA of $L1_0$-FePt, \cite{Solovyev_RSGreens_FePt,Ueda_spinflip_HXPES_FePt} 
some first-principles calculations show that the MCA energy from the Fe site can be larger than that from the Pt site and that the large SOC of Pt only indirectly participates through the Fe $3d$-Pt $5d$ band hybridization. \cite{Burkert_FePt_FeDominant, Zhu_MCAtheory_JPCM2013} 
Thus, consensus has not been reached yet about the microscopic origin of the PMA of $L1_0$-FePt. 

In order to elucidate the origin of the PMA in $L1_0$-FePt, 
it is essential to experimentally probe the orbital asphericity as well as $m_\text{orb}$ of the Fe $3d$ and Pt $5d$ electrons. 
XMCD is an element-specific technique which can separately probe the orbital and spin magnetic moments 
of ferromagnetic materials by using XMCD sum rules. \cite{OrbSum, SpinSum} 
Moreover, it is known that the deduced spin magnetic moment from the XMCD sum rule, 
hereafter called as `effective' spin magnetic moment ($m_\text{spin}^\text{eff}$), 
contains an additional term called `magnetic dipole moment'  ($m_\text{T}$), 
which is directly connected to the orbital asphericity of the spin-polarized electrons. \cite{SpinSum, StohrKonig, DurrTXMCD, Stohr_JMMM1999, Durr_Science97}
It is known that the magnetic dipole moment $m_\text{T}$ can be separated from the spin magnetic moment ($m_\text{spin}$) 
by measuring the magnetic-field angle dependence of the XMCD spectra. \cite{DurrTXMCD, Stohr_JMMM1999, Durr_Science97, vanderLaanPRL10, AngleDep_shibata, AngleDep_sakamoto}
In the present study,  
in order to reveal the orbital asphericity of the Fe $3d$ electrons, we have performed 
angle-dependent XMCD (AD-XMCD) experiments at the Fe $L_{2,3}$ edge of FePt thin films 
with different degrees of $L1_0$ order and hence different MCA energies. 
From the magnetic field-angle dependence of $m_\text{spin}^\text{eff}$, 
it has been shown that 
the spin-polarized electrons of Fe 
preferentially occupy orbitals elongated along the $z$ axis (the out-of-plane direction). 
The deduced MCA energy due to the orbital asphericity is negative, 
and its magnitude is larger than that due to the OMA. 
This means that the Fe sites favor the in-plane magnetic easy axis 
and that Pt should play a dominant role in the PMA of $L1_0$-ordered FePt thin films. 

%%%%%%%%%%%%%%%%% FIGURE 1 (Geometry & Angle-dependent spectra)
\begin{figure*}[t]%[H]
\begin{center}
\includegraphics[width=17cm]{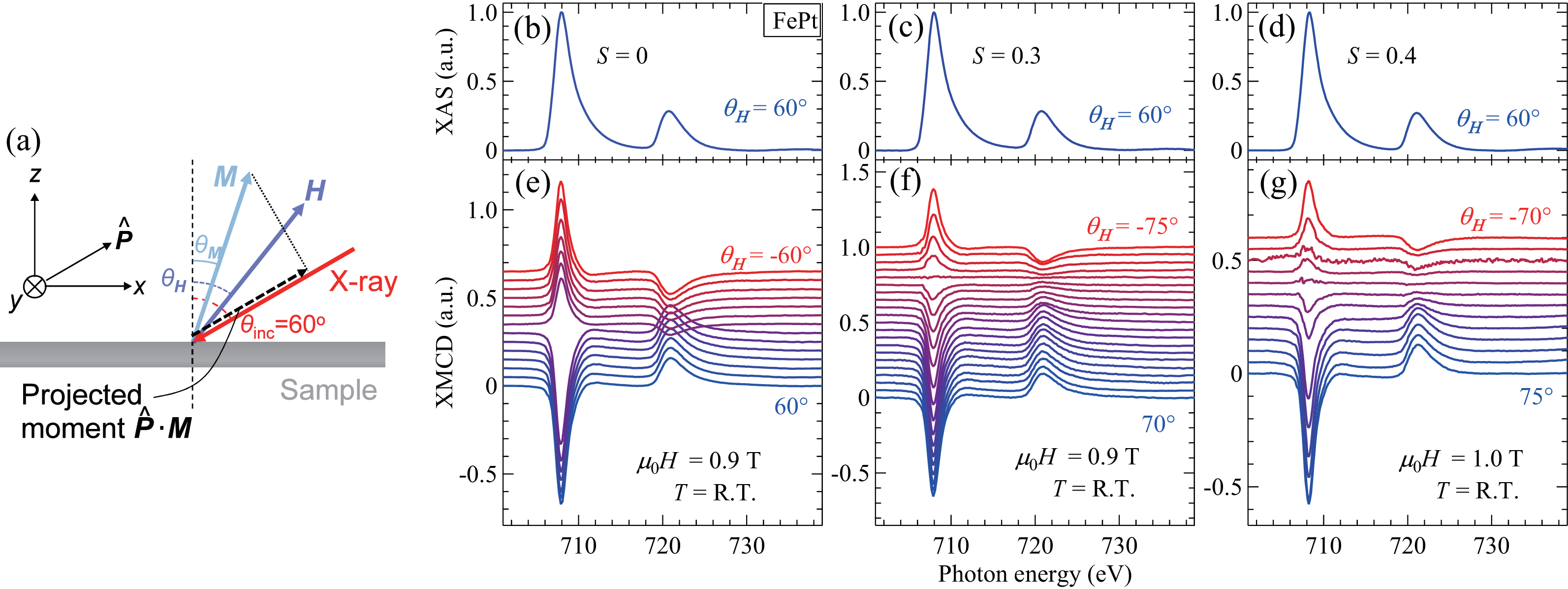}
\caption{Fe $L_{2,3}$-edge x-ray absorption spectroscopy (XAS) and angle-dependent x-ray magnetic circular dichroism (AD-XMCD) measurements. (a) Schematic illustration of the experimental geometry. 
The angles of the incident x ray ($\theta_\text{inc}$),  the external magnetic field ($\theta_{\bm{H}}$) and the magnetization ($\theta_{\bm{M}}$) are defined with respect to the sample normal. 
$\theta_\text{inc}$ is fixed at $60^\circ$ in the present measurements. 
$\theta_{\bm{H}}$ can be varied in the range of $-{90}^\circ \leq \theta_{\bm{H}} \leq {90}^\circ$. 
Note that the external magnetic field $\bm{H}$ and the magnetization $\bm{M}$ are not always parallel 
in the presence of magnetic anisotropy. 
(b--d) XAS and (e--g) AD-XMCD spectra of the FePt thin films 
with $L1_0$ order degrees $S=0$ (b,e), $S=0.3$ (c,f) and $S=0.4$ (d,g). 
Linear and smoothed two-step backgrounds have been subtracted from the raw XAS spectra. 
All the XAS spectra have been normalized to the height of the $L_3$ peak at $\sim 708\ \text{eV}$. 
$\theta_{\bm{H}}=60^\circ$ corresponds to the geometry where the incident x ray and the magnetic field are parallel 
[See Panel (a)]. 
XMCD intensity at the $L_3$ edge is proportional to the projected magnetic moment $\hat{\bm{P}}\cdot\bm{M}$ 
shown in Panel (a). 
}
\label{fig:ADspectra}
\end{center}
\end{figure*}
%%%%%%%%%%%%%%%%% FIGURE 1 UNTIL HERE

%%%%%%%%%%%%%%%%% Table 1 (Sample preparation conditions)
\begin{table}[b]%[H]
\begin{center}
\caption{
Sample-preparation conditions, the degree of the $L1_0$ order $S$, 
magnetic anisotropy energy due to the demagnetizing field (shape anisotropy) $\frac{\mu_0}{2} M_\text{s}^2$, 
and the magnetocrystalline anisotropy (MCA) energy deduced from magnetization curves $K_\text{u}^{M\text{-}H}$.
We note that 
$K_\text{u}^{M\text{-}H}$ is the sum of the shape anisotropy energy and the effective magnetic anisotropy energy $K_\text{eff}$ calculated directly from the $M$-$H$ curves. 
}
\label{table:sample_prep}
\vspace*{0.3cm}
\begin{ruledtabular}
\begin{tabular}{ccccc}
%\hline\hline
 \shortstack{Deposition \\temperature $T_\text{S}$ ($^\circ\mathrm{C}$)} & \shortstack{Annealing \\temperature $T_A$ ($^\circ\mathrm{C}$)} & \shortstack{$L1_0$ order\\ degree $S$} & \shortstack{Shape anisotropy energy \\ $\frac{\mu_0}{2} M_\text{s}^2$} & \shortstack{$K_\text{u}^{M\text{-}H}$\\(MJ/m$^3$)}\\
\hline
300 & 350 & 0.4 & 0.80 & 1.3\\

300 & -- & 0.3 & 0.71 & 0.7\\

RT & -- & 0 & 0.51 & 0.3\\
%\hline\hline
\end{tabular}
\end{ruledtabular}
\end{center}
\end{table}
%%%%%%%%%%%%%%%%% Table 1 UNTIL HERE

%%%%%%%%%%%%% SAMPLE PREPARATION
FePt thin films were grown on MgO (100) substrates by the ultrahigh vacuum magnetron-sputtering method. 
The details of the sample preparation methods are described in the previous report. \cite{FePt_Seki_fund} 
The stacked structure of the sample was MgO substrate/Fe (1 nm)/Au (30 nm)/FePt (20 nm)/Au (2 nm). 
The Au 2 nm layer at the top was deposited as a capping layer in order to prevent the sample from oxidization. 
The 30-nm-thick Au layer was deposited as a buffer layer to ensure that the lattice constant of FePt is not influenced 
by that of the substrate, and the 1-nm-thick Fe layer is the seed layer. 
One can control the degree of the $L1_0$ order $S$ 
(the definition of which is described elsewhere \cite{def_of_S1, def_of_S2}) 
by changing the substrate temperature during deposition $T_\text{S}$ 
or the subsequent annealing temperature after the deposition $T_\text{A}$. \cite{FePt_Seki_fund} 
Table\ \ref{table:sample_prep} summarizes the sample-preparation conditions, 
the degree of the $L1_0$ order $S$ 
deduced from the intensity of the (001) x-ray diffraction peak, 
and the MCA energy deduced from magnetization measurements $K_\text{u}^{M\text{-}H}$ 
(see Supplementary Materials for the x-ray diffraction profiles and the magnetization curves). 
$K_\text{u}^{M\text{-}H}$ was calculated as the sum of the effective magnetic anisotropy energy $K_\text{eff}$ and the general shape anisotropy energy for thin films, $\frac{\mu_0}{2} M_\text{s}^2$, where $M_\text{s}$ is the saturation magnetization. $K_\text{eff}$ was calculated from the area surrounded by the in-plane and out-of-plane magnetization curves and $M_\text{s}$ was defined as the average magnetization in the range of $3\ \text{T} \leq |\mu_0 H|\leq 5\ \text{T}$. 

%%%%%%%%%%%%%%% XMCD MEASUREMENT CONDITIONS
The Fe $L_{2,3}$-edge x-ray absorption spectroscopy (XAS) and AD-XMCD measurements 
were performed at helical undulator beamline BL-16A of Photon Factory, 
High Energy Accelerator Research Organization (KEK-PF). 
Figure \ref{fig:ADspectra}(a) is the schematic illustration of the measurement geometry. 
The x-ray incident angle $\theta_\text{inc}$ was fixed at $60^\circ$ from the sample normal 
and the magnetic field direction relative to the sample normal $\theta_{\bm{H}}$ was varied in the range of $-90^\circ \leq \theta_{\bm{H}} \leq +90^\circ$ 
(where the sample normal is defined as $0^\circ$) using two orthogonal pairs of superconducting magnets. \cite{Vector_Furuse, AngleDep_shibata,AngleDep_sakamoto,AngleDep_nonaka} 
%Here, the sample normal is defined as $\theta = {0}^\circ$, and the x ray incident direction to be $+{60}^\circ$ (Fig.\ \ref{fig:setup_FePt}).
From the different angular dependencies between the magnetic dipole moment $m_\text{T}$ and the spin magnetic moment $m_\text{spin}$, one can deduce $m_\text{T}$ by AD-XMCD. \cite{StohrKonig,DurrTXMCD,AngleDep_shibata,AngleDep_sakamoto}
The maximum magnitude of the applied magnetic field was 1 T. 
In order for the potential experimental artifacts to be cancelled out, the XMCD measurements were performed for the positive and negative (reversed) magnetic fields at each magnetic field angle $\theta_{\bm{H}}$ and the two obtained XMCD spectra were averaged. 
All the measurements were performed at room temperature. 
Absorption signals were collected in the total electron-yield method. 
On the application of the XMCD sum rules, we first subtracted the polygonal line bent at the Fe $L_3$ XAS peak and then a two-step white line background composed of arctangent functions.  
On the application of the spin sum rule \cite{SpinSum}, the boundary between the Fe $L_3$ and $L_2$ edges was set to be $h\nu = 718$ eV and 
the hole number $n_h$ was assumed to be 3.4 based on a previous band-structure calculation.\cite{Antoniak_FePt_NanoP}
Details of the sum rule analysis have been described in one of our previous report. \cite{Shibata_FexTiS2} 
We note that the arbitrariness of these procedures on applying the XMCD sum rules can result in systematic errors of $\sim \pm (10-15)\% $ in the deduced magnetic moments. \cite{Shibata_FexTiS2} 

\begin{figure}%[H]
\begin{center}
\includegraphics[width=8cm]{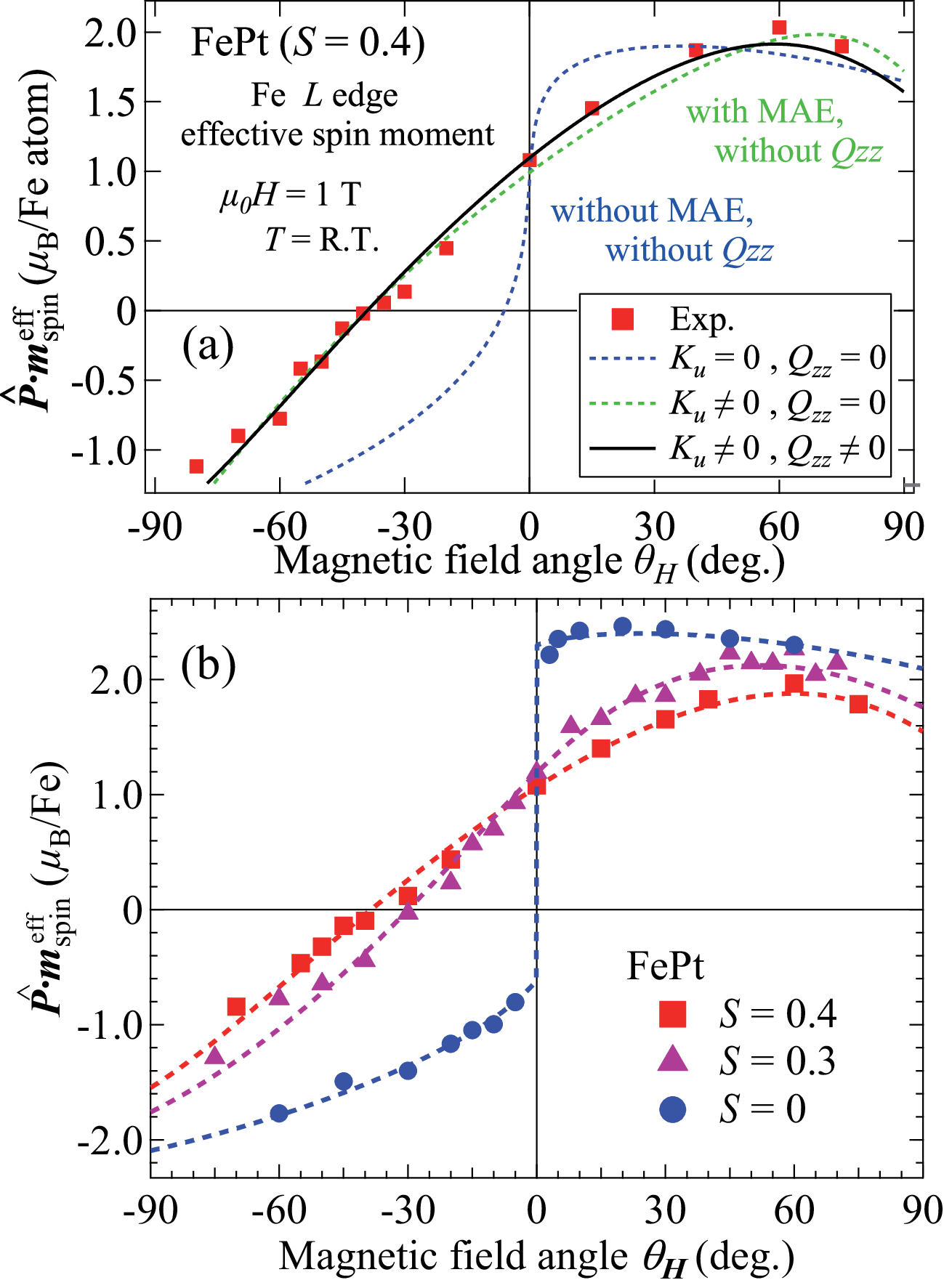}
\caption{
Magnetic field angle dependence of effective spin magnetic moment projected on the light axis 
($\hat{\bm{P}}\cdot\bm{m}_\text{spin}^\text{eff}$). 
(a) Experimental data for the film with $S=0.4$ (red square) and simulated angular dependencies 
based on the Stoner-Wohlfarth model without magnetocrystalline anisotropy 
(MCA) and $\braket{ Q_{zz} }$ (blue dotted curve), 
with MCA but without $\braket{ Q_{zz} }$ (green dotted curve), 
and with both MCA and $\braket{ Q_{zz} }$ (black curve). 
(b) Experimental and simulated angular dependencies of $\hat{\bm{P}}\cdot\bm{m}_\text{spin}^\text{eff}$ 
for various $L1_0$ order degrees $S$. 
Both $K_\text{u}$ and $\braket{ Q_{zz} }$ are incorporated in the simulation. 
} 
\label{fig:AD_spin}
\end{center}
\end{figure}

%%%%%%%%%% RESULTS 1: ANGLE-DEPENDENT SPECTRA 
Figures\ \ref{fig:ADspectra}(b--d) and \ref{fig:ADspectra}(e--g), respectively, show the XAS %x-ray absorption spectroscopy (XAS) 
spectra and AD-XMCD spectra at the Fe $L_{\rm 2,3}$ edge of the {\sl L}1$_{0}$-FePt thin films with $S$ = 0, 0.3, and 0.4. 
All the XAS spectra show similar line shapes to those of metallic Fe, \cite{CTChen_FeXAS,FePt_Ikeda} 
and no oxidation features are observed. 
With varying magnetic field angle $\theta_{\bm{H}}$, the XMCD abruptly reverses the sign at $\theta_{\bm{H}}=0^\circ$ for $S=0$. 
In contrast, for $S=0.3$ and 0.4, the XMCD spectra gradually change the intensity and sign, as shown in Fig.\ \ref{fig:AD_spin}(b). 
This corresponds to the change in the magnetization angle ($\theta_{\bm{M}}$) with varying $\theta_{\bm{H}}$, 
since the XMCD intensity is generally proportional to the projected magnetic moment onto the light axis [see Fig.\ \ref{fig:ADspectra}(a)]. 
It should be noted that the external magnetic field $\bm{H}$ and the magnetization $\bm{M}$ are not necessarily parallel 
(i.e. $\theta_{\bm{H}} \neq \theta_{\bm{M}}$) in the presence of magnetic anisotropy, 
because $\bm{M}$ tends to be oriented along the magnetic easy axis. 
Based on this principle, one can deduce the MCA energy ($K_\text{u}$) of each film from the angular dependencies of the XMCD intensity, 
as shown below. 

%%%%%%%%%%%%% RESULTS 2: ANGLE-DEPENDENCE OF M_SPIN^EFF AND CURVE FITTING

In order to analyze the angular dependence of the XMCD intensity more quantitatively, 
we have deduced the spin magnetic moment by applying the XMCD spin sum rules \cite{SpinSum} 
to each spectrum in Figs.\ \ref{fig:ADspectra}(e--g). 
By using the XMCD spin sum rule, 
the `effective' spin magnetic moment (per Fe ion) 
$\bm{m}_\text{spin}^\text{eff} \equiv \bm{m}_\text{spin} + (7/2)\bm{m}_\text{T}$
projected onto the x ray propagation vector ($\hat{\bm{P}}$), 
i.e., $\hat{\bm{P}}\cdot \bm{m}_\text{spin}^\text{eff}$, 
is deduced \cite{SpinSum}. 
Here, the magnetic dipole moment $\bm{m}_\text{T}$ is defined by 
$\bm{m}_\text{T} \equiv -g \mu_{\text{B}} \langle \bm{S}-3(\bm{r}\cdot\bm{S})/r^2 \rangle = -g \mu_{\text{B}} \langle \bm{Q}\cdot \bm{S} \rangle$, 
where $\bm{r}$, $\bm{S}$, and $Q_{ij} \equiv \delta_{ij}-(r_i r_j)/r^2$ are the position operator, spin angular momentum operator, and electric quadrupole tensor operator, respectively. \cite{SpinSum, StohrKonig, DurrTXMCD, Stohr_JMMM1999} 
By definition, $\bm{m}_\text{T}$ is a vector related to both the spin magnetic moment and the asphericity of the charge density, i.e., elongation or shrinkage of the electron orbitals. 
Since $\hat{\bm{P}} \cdot \bm{m}_\text{spin}$ and $\hat{\bm{P}} \cdot \bm{m}_\text{T}$ 
show different angular dependencies, one can extract the latter term by measuring the angular dependence of $\hat{\bm{P}}\cdot \bm{m}_\text{spin}^\text{eff}$, and thus information about the orbital asphericity is obtained. 
\cite{DurrTXMCD, Durr_Science97, vanderLaanPRL10, AngleDep_shibata, AngleDep_sakamoto}

\begin{figure}%[H]
\begin{center}	
\includegraphics[width=7cm]{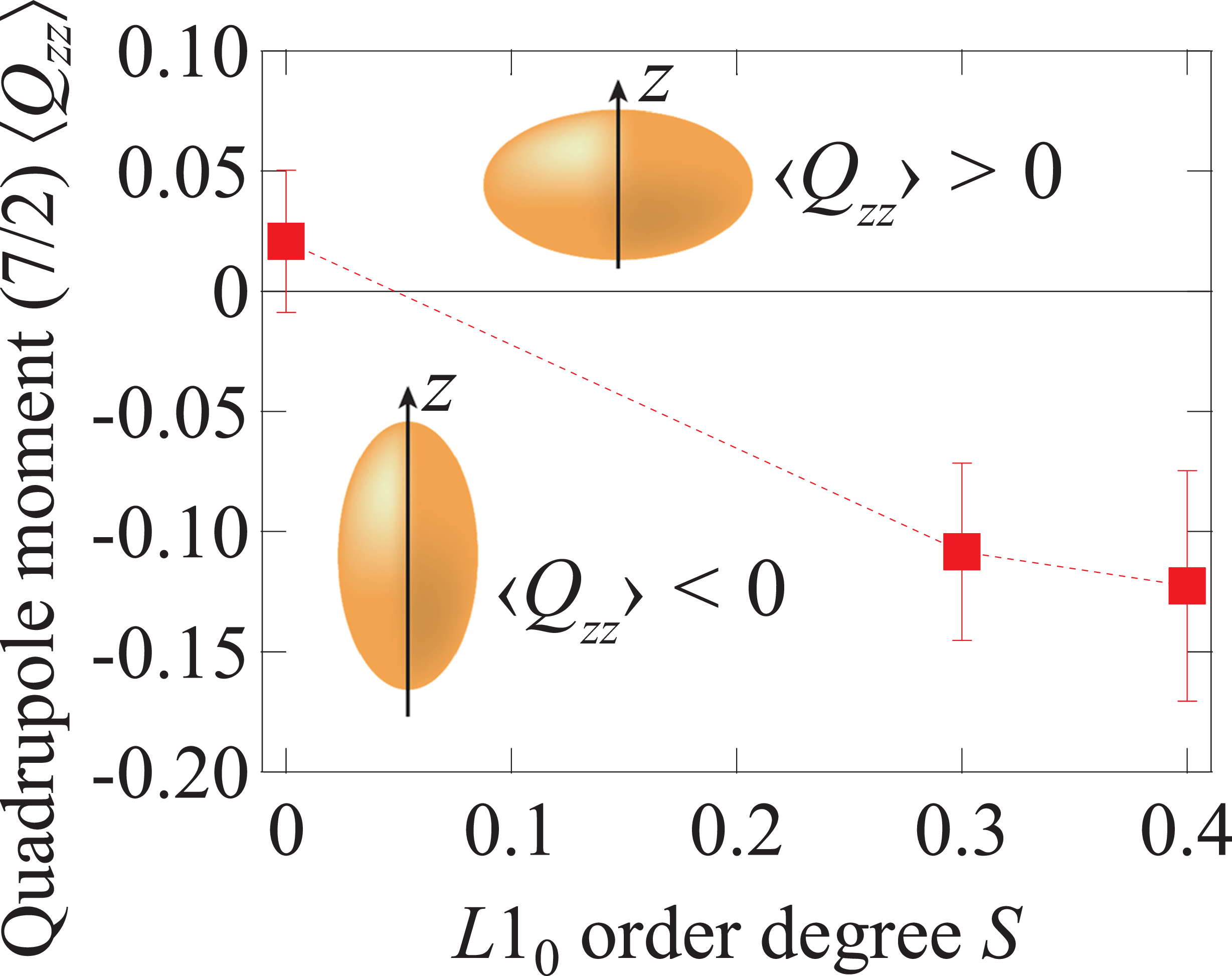}
\caption{ Electric quadrupole moment $(7/2)\braket{ Q_{zz} }$ of Fe as a function of $L1_0$ order degree $S$. 
Error bars represent the standard deviation deduced from the fit in Fig.\ \ref{fig:AD_spin}. 
The ellipsoids schematically describe the charge densities for $\braket{ Q_{zz} }>0$ and $\braket{ Q_{zz} }<0$.  
 } 
\label{fig:Qzz_ver_S}
\end{center}
\end{figure}

The $\theta_{\bm{H}}$ dependence of $\hat{\bm{P}}\cdot \bm{m}_\text{spin}^\text{eff}$ for the film with $S$ = 0.4 
obtained by the sum rule is shown in Fig.\ \ref{fig:AD_spin}(a), 
and is analyzed based on the Stoner-Wohlfarth model, \cite{SWmodel_original}
in which one assumes the uniaxial MCA and single magnetic domain. 
Under these assumptions, the magnetostatic energy density $E_\text{m}$ is given by: 
\begin{equation}
E_\text{m} = - \mu_0 MH\cos(\theta_{\bm{M}}-\theta_{\bm{H}}) + \frac{\mu_0}{2}M^2 \cos^2 \theta_{\bm{M}} - K_\text{u} \cos^2 \theta_{\bm{M}}, 
\label{EnergyEq}
\end{equation}
where $M$ is the total magnetization (the sum of the spin and orbital magnetic moments of Fe and Pt). 
The three terms in Eq.\ (\ref{EnergyEq}) represent the Zeeman energy, 
the general magnetic shape anisotropy for thin films, 
and the MCA energy. 
By minimizing $E_m$ with respect to $\theta_{\bm{M}}$, one can deduce the value of $\theta_{\bm{M}}$ for each angle $\theta_{\bm{H}}$, 
and thus the effective spin magnetic moment can be deduced by the following equation:  \cite{AngleDep_shibata,AngleDep_sakamoto,AngleDep_nonaka} 
\begin{eqnarray}
& & \hat{\bm{P}} \cdot \bm{m}_{\text{spin}} + (7/2) \hat{\bm{P}} \cdot \bm{m}_{\text{T}} \nonumber \\
& \simeq & m_{\text{spin}} \cos (\theta _{\bm{M}} - \theta _{\text{inc}}) + (7/4) \langle Q_{zz} \rangle m_{\text{spin}} (2 \cos \theta _{\bm{M}} \cos \theta _{\text{inc}} - \sin \theta _{\bm{M}} \sin \theta _{\text{inc}}), 
\label{angledep}
\end{eqnarray}
where $\theta _{\text{inc}} = 60^{\circ}$ is the x-ray incidence angle. 
In the simulation, we have assumed that the angular dependence of $m_\text{spin}+m_\text{orb}$ is negligible 
and that the ratio of $m_\text{spin}^\text{Pt}$ to $m_\text{spin}^\text{Fe}$ is constant 
($m_\text{spin}^\text{Pt} / m_\text{spin}^\text{Fe} \simeq 0.11$)\cite{FePt_Ikeda}. 
The curves in Fig.\ \ref{fig:AD_spin}(a) represent thus simulated angular dependencies 
under various assumptions: (i) both MCA ($K_\text{u}$) and quadrupolar moment $\braket{ Q_{zz} }$ are absent, (ii) MCA is incorporated but $\braket{ Q_{zz} }$ is absent, and (iii) both MCA and $\braket{ Q_{zz} }$ are taken into account. 
Simulation (i) clearly deviates from the experimental data, showing that one needs to incorporate MCA in order to reproduce the experimental angular dependence. 
Comparing simulations (ii) and (iii), the latter (simulation with $\braket{ Q_{zz} }$) better agrees with the experimental data. 
Figure \ref{fig:AD_spin}(b) shows the experimental and simulated angular dependencies of the effective spin magnetic moment for all the samples based on simulation. 
All the data could be well reproduced based on the Stoner-Wohlfarth model, suggesting that the assumption of the magnetic single domain is valid.
The fitted values of $m_\text{spin}$, $K_\text{u}$, and $\braket{ Q_{zz}}$
are summarized in Table \ref{table:fitted_variables_FePt}. 
The MCA energy $K_\text{u}$ increases with $S$, consistent with the magnetization measurements (Fig.\ S2 of Supplementary Materials) and the previous magnetization studies \cite{K1K2_FePt_Okamoto,FePt_Seki_fund}. 
One can see that $\braket{ Q_{zz}}$ is negative for the films which exhibit PMA ($S=0.3, 0.4$) 
and that the magnitude of $|\braket{ Q_{zz}}|$ increases with $S$ and $K_\text{u}$. 
Figure \ref{fig:Qzz_ver_S} summarizes the $S$ dependence of $\braket{ Q_{zz}}$. 
The negative $\braket{ Q_{zz}}$ values show that the charge density in Fe ions is elongated along the out-of-plane direction, 
as shown in the bottom of Fig.\ \ref{fig:Qzz_ver_S}. 
We note that the decrease of $m_\text{spin}$ with $S$ in Table \ref{table:fitted_variables_FePt} is consistent with the previous experimental study\cite{FePt_Ikeda} and the theoretical calculations based on the coherent potential approximation. \cite{Staunton_FePt_CPA_spin}

\begin{table}%[H]
\begin{center}
\caption{
Fitting parameters in the Stoner-Wohlfarth model for each sample. 
}
\label{table:fitted_variables_FePt}
\begin{ruledtabular}
\begin{tabular}{cccc}
$S$& $m_{\rm spin}$ ($\mu_{\rm B}$/Fe) & $K_\text{u}$ (MJ/m$^3$)  &$(7/2)\braket{ Q_{zz} }$  \\
\hline
0.4 & 1.9 & 4.1 & $-0.12 \pm 0.05$\\

0.3 & 2.2 & 3.9 & $-0.11 \pm 0.04$\\

   0 & 2.4 & -0.3 & $0.02 \pm 0.03$\\
\end{tabular}
\end{ruledtabular}
\end{center}
\end{table}

%%%%%%%%%%%%%%%% RESULTS 3: Contributions of spin-conservation and spin-flip terms to MAE %%%%%%%%%%%%%

\begin{figure}%[H]
\begin{center}
\includegraphics[width=8cm]{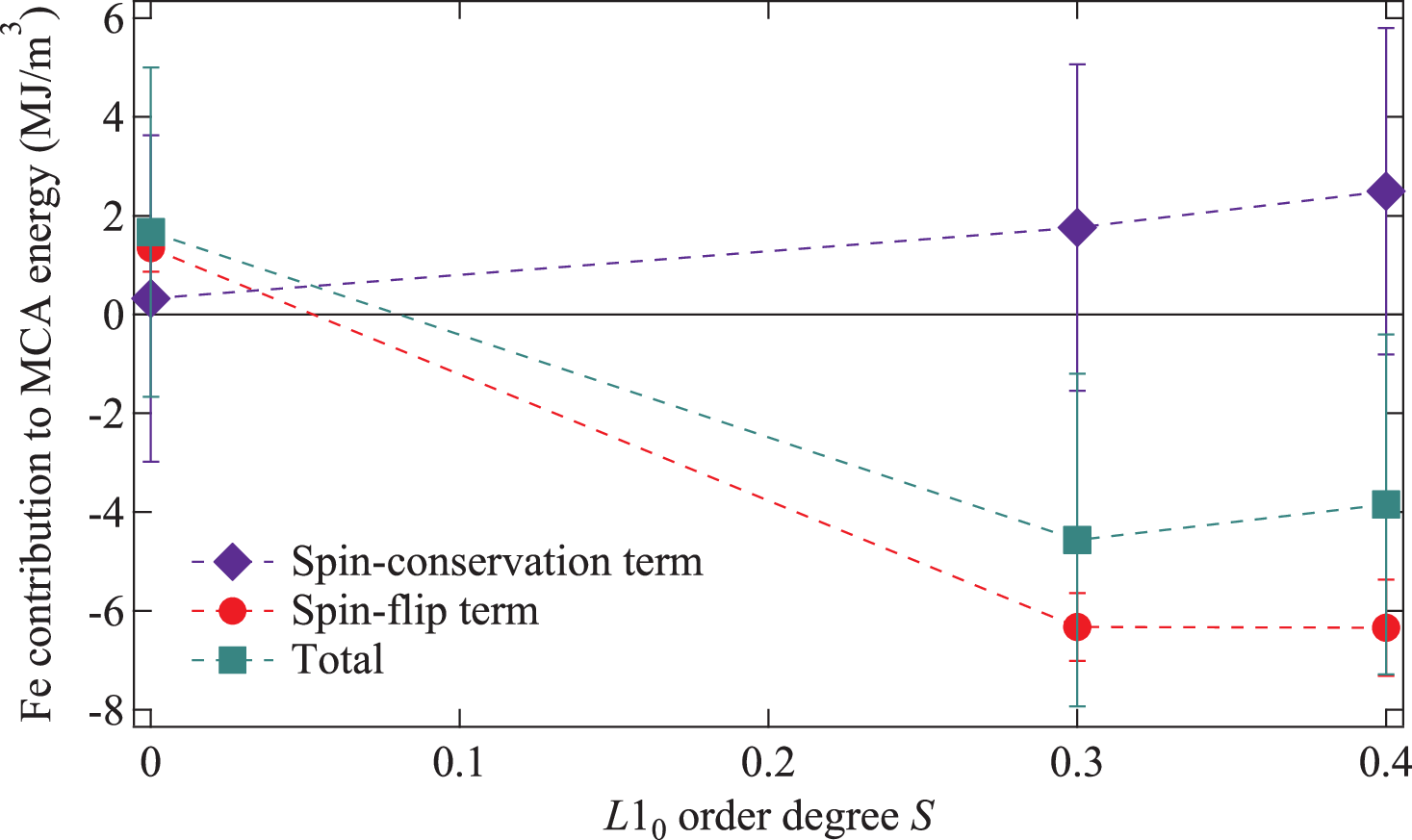}
\caption{
Contribution to the MCA energy originating from the orbital moment anisotropy (spin-conservation term, blue) 
and the charge-density anisotropy (spin-flip term, red). 
The sum of the two terms (total MCA energy) is also shown (green). 
} 
\label{fig:MAE_XMCD}
\end{center}
\end{figure}

According to the previous theoretical studies \cite{Wang_spinflip,vanderLaan}, 
the MCA energy ($K_\text{u}$) can be expressed as the sum of two terms 
originating from spin-conserved and spin-flip virtual excitations in the second-order perturbation theory 
with respect to SOC: 
specifically, in the well-known formula of second-order perturbation theory \cite{Brunoeq,Wang_spinflip,vanderLaan}
\begin{eqnarray*}
E_2 &=&
 \sum_{\rm ex}\frac{\braket{{\rm gr}|H_{\rm SO}|{\rm ex}}\braket{{\rm ex}|H_{\rm SO}|{\rm gr}}}{E_{\rm gr} - E_{\rm ex}}, \\
\ket{\rm ex} & = & c^{\dagger}_{n_2,\sigma_2}c_{n_1,\sigma_1} \ket{\rm gr} \qquad (\sigma_1, \sigma_2 = \pm 1/2),
\label{perturb}
\end{eqnarray*}
the terms with $\sigma_2 = \sigma_1$ and $\sigma_2 = -\sigma_1$ are referred to as spin-conserved and spin-flip terms, respectively. 
Under the assumption that the majority-spin band is fully occupied,  
the explicit form of $K_\text{u}$ within the perturbation theory is: \cite{Wang_spinflip,vanderLaan} 
\begin{eqnarray}
K_\text{u} 
&=& E_2^{\parallel} - E_2^{\perp} \nonumber \\
&=& \frac{\zeta}{4\mu_\text{B}} \left(m_\text{orb}^{\perp} - m_\text{orb}^{\parallel} \right) 
+ \frac{21}{4\mu_\text{B}}\frac{\zeta^2}{\Delta E_\text{ex}} \left( m_\text{T}^{\perp} - m_\text{T}^{\parallel} \right) \nonumber \\
&=&\frac{\zeta}{4\mu_\text{B}} \left(m_\text{orb}^{\perp} - m_\text{orb}^{\parallel} \right) 
+ \frac{21}{4\mu_\text{B}}\frac{\zeta^2}{\Delta E_\text{ex}} m_\text{spin}\braket{Q_{zz}},
 \label{MAEEq}
\end{eqnarray}
where $\zeta$ is the SOC constant and $\Delta E_\text{ex}$ is the exchange splitting between the majority- and minority-spin bands. 
The first term, originating from spin-conserved excitations, is proportional to the orbital moment anisotropy (OMA) between the film out-of-plane and in-plane directions. 
The second term, originating from spin-flip excitations, is proportional to the anisotropic charge distribution $\braket{Q_{zz}}$. 
Using the OMA 
and the $\braket{Q_{zz}}$ values deduced in the present study (Table \ref{table:fitted_variables_FePt} and Fig.\ \ref{fig:Qzz_ver_S}), 
we have estimated the contributions of Fe to the MCA of FePt thin films 
and have plotted them in Fig.\ \ref{fig:MAE_XMCD} as functions of the $L1_0$ order degree $S$.
Here, $\zeta$ and $\Delta E_\text{ex}$ of Fe have been assumed to be 0.054 eV and 3 eV, respectively \cite{vanderLaan,Kota_FePtcalc}. 
The OMA is positive and gives a small positive contribution to the PMA.
In contrast, the negative spin-flip term of Fe favors the in-plane magnetic easy axis.
% which is opposite to the spin-conservation term. 
Besides, its magnitude is significantly larger than that of the spin-conservation term.
Therefore, the sum of the spin-conservation term and the spin-flip term is also negative with increasing $S$, as shown in Fig. \ref{fig:MAE_XMCD}. 

%%%%%%%%%%%%%%%%%%%%%%%%%%%%%%%%%%% DISCUSSION %%%%%%%%%%%%%%%%%%%%%%%%%%%%5
As stated above, there are discrepancies on the origin of the PMA of $L1_0$ FePt thin films among the different theoretical and experimental studies. Some studies suggest that Fe has a dominant contribution to the PMA \cite{Burkert_FePt_FeDominant, Zhu_MCAtheory_JPCM2013}, while others suggest that Pt contribution is more significant \cite{Solovyev_RSGreens_FePt,Ueda_spinflip_HXPES_FePt}. 
The present XMCD results show that the overall MCA energy from the Fe ions is negative, which rules out the former standpoint. The negative MCA energy due to the orbital asphericity ($m_{\rm T}$), revealed by the present AD-XMCD measurements, dominates over the positive contribution due to the OMA of Fe. 

The present results also indicate that the PMA of $L1_0$ FePt thin films mainly originates from the Pt sites. Our previous XMCD study at the Pt $L_{2,3}$ edge \cite{FePt_Ikeda} shows that the OMA of Pt is negative ($m_\text{orb}^{\perp} < m_\text{orb}^{\parallel}$), suggesting that the OMA favors the in-plane magnetic easy axis. 
The study also suggests that $m_\text{spin}^\text{eff}$ of Pt along the out-of-plane direction is slightly larger than that along the in-plane direction for highly ordered samples, which might be explained by the increase of $m_\text{T}^{\perp} - m_\text{T}^{\parallel}$ with increasing $S$. \cite{FePt_Ikeda} 
According to Eq.\ (\ref{MAEEq}), this has a positive contribution to PMA. 
Therefore, it follows that the PMA of the $L1_0$ FePt thin films is primarily due to the orbital asphericity of the Pt sites. 
These results are consistent with the theoretical calculation using the Green's function method \cite{Solovyev_RSGreens_FePt}. 
To verify this hypothesis, AD-XMCD experiments at the Pt $L_{2,3}$ edge using hard x rays 
could provide a direct evidence, which remain to be performed in the future. 

%%%%%%%%%%%%%%%%%%%%% SUMMARY %%%%%%%%%%%%%%%%%%%%%%
In summary, we have performed AD-XMCD measurements of FePt thin films with various $L1_0$ order degrees $S$. 
The MCA energy $K_\text{u}$ and the electric quadrupole moment $\braket{ Q_{zz} }$, i.e. the charge-density anisotropy, have been quantitatively deduced from the angular dependence of XMCD. 
The deduced $\braket{ Q_{zz} }$ is negative for the films with larger $S$, suggesting that the Fe 3$d$ orbitals elongated along the $z$ direction are preferentially occupied in the $L1_0$-ordered films. 
The MCA energy due to this charge-density anisotropy of Fe ions is negative and dominates over the positive contribution by OMA, 
consistent with the previous first-principles calculation \cite{Solovyev_RSGreens_FePt}. 
Considering that the OMA of Pt also favors the in-plane magnetic easy axis, \cite{FePt_Ikeda} one has to take into account the charge-density anisotropy of Pt in order to explain the large PMA of $L1_0$-ordered FePt thin films.  

\section*{Supplementary Material}
See supplementary material for the X-ray diffraction data and the magnetization curves of the grown samples. 

\begin{acknowledgments}
We would like to thank Kenta Amemiya and Masako Suzuki-Sakamaki for valuable technical support at KEK-PF. We would also like to thank Iwao Matsuda for valuable advice and enlightening discussion. This work was supported by Grants-in-Aid for Scientific Research from JSPS (grant Nos.~15H02109,  15K17696, 20K14416, and 22K03535), 
MEXT Initiative to Establish Next-generation Novel Integrated Circuits Centers (X-NICS) Grant Number JPJ011438, 
and the National Science and Technology Council of Taiwan under Grant No. 113-2112-M-007-033. The experiment was performed at BL-16A of KEK-PF with the approval of the Photon Factory Program Advisory Committee (proposal Nos.~2016G066, and 2016S2-005). A.F. is an adjunct member of Center for Spintronics Research Network (CSRN), the University of Tokyo, under Spintronics Research Network of Japan (Spin-RNJ), and acknowledges support from the Yushan Fellow Program under the Ministry of Education of Taiwan. T.S., S.S. and K.T. are members of CSRN, Tohoku University, under Spin-RNJ. 
\end{acknowledgments}

\section*{AUTHOR DECLARATIONS}
\subsection*{Conflict of Interest}
The authors declare no conflicts of interest.

\subsection*{Author Contributions}
\textbf{Goro Shibata:} Formal analysis (equal); Funding acquisition (supporting); Investigation (equal); Methodology (equal); Software (equal); Visualization (equal); Writing – original draft (equal)
\textbf{Keisuke Ikeda:} Formal analysis (equal); Investigation (equal); Methodology (equal); Software (equal); Visualization (equal); Writing – original draft (equal)
\textbf{Takeshi Seki:} Resources (lead); Funding acquisition (supporting); Writing -- review and editing (equal)
\textbf{Shoya Sakamoto:} Investigation (supporting); Writing -- review and editing (equal)
\textbf{Yosuke Nonaka:} Investigation (supporting)
\textbf{Zhendong Chi:} Investigation (supporting)
\textbf{Yuxuan Wan:} Investigation (supporting)
\textbf{Masahiro Suzuki:} Investigation (supporting)
\textbf{Tsuneharu Koide:} Methodology (supporting); Resources (supporting); Supervision (equal); Writing -- review and editing (equal) 
\textbf{Hiroki Wadati:} Supervision (equal) 
\textbf{Koki Takanashi:} Conceptualization (equal); Project administration (equal); Supervision (equal); Writing -- review and editing (equal)
\textbf{Atsushi Fujimori:} Conceptualization (equal); Funding acquisition (lead); Project administration (equal); Supervision (equal); Writing – original draft (equal)

\section*{AVAILABILITY OF DATA}
The data that support the findings of this study are available from the corresponding author upon reasonable request.

\bibliography{ref}

\setcounter{figure}{0}
\renewcommand{\thefigure}{S\arabic{figure}}

\newpage

\section*{Supplementary Material}

\begin{figure}[h]
\includegraphics[width=16cm]{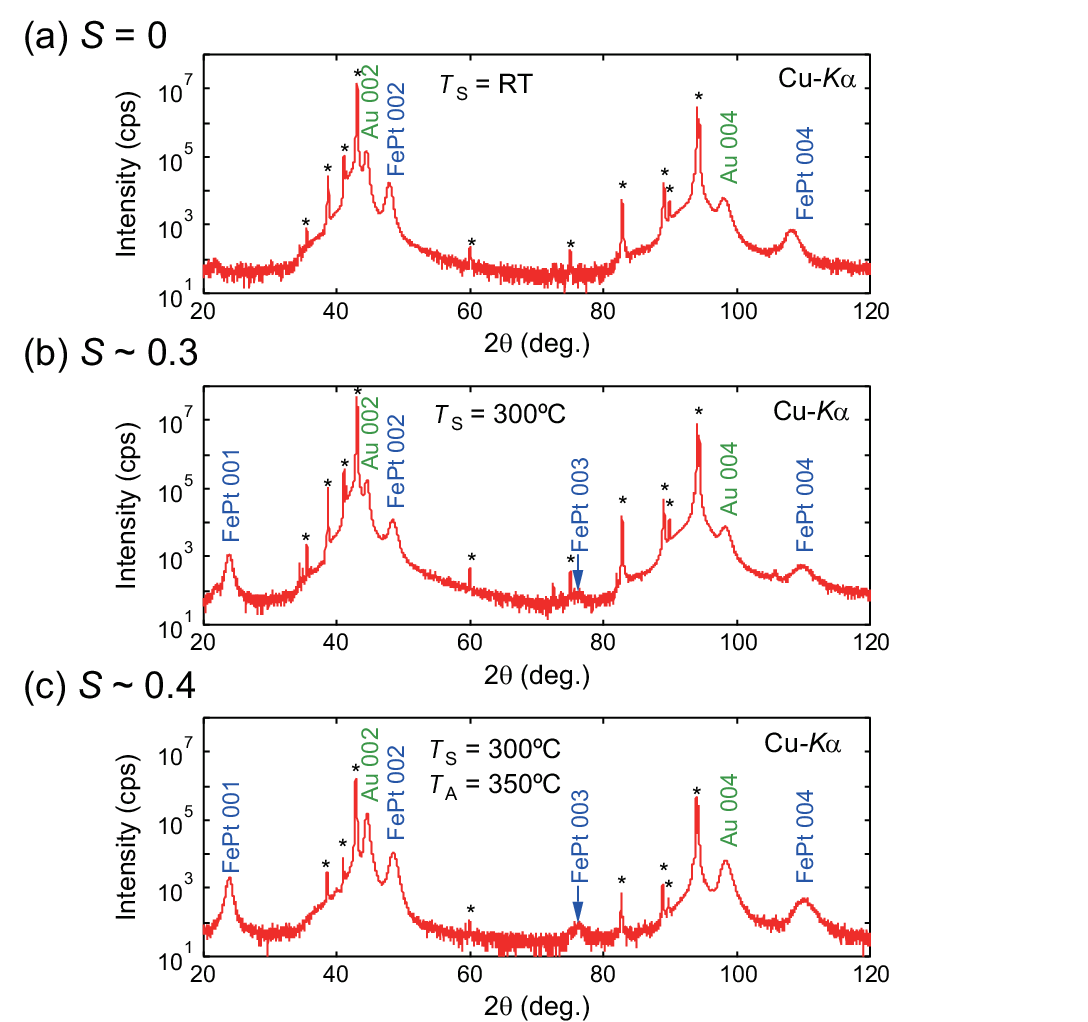}
\caption{X-ray diffraction (XRD) patterns of the grown $L1_0$ FePt films with $L1_0$ order degrees $S=0$, $S \sim 0.3$, and $S \sim 0.4$. $T_\text{S}$ and $T_\text{A}$ represent the substrate temperature during the growth and the post-annealing temperature, respectively. 
\label{XRD}}
\end{figure}

\newpage

\begin{figure}
\includegraphics[width=16cm]{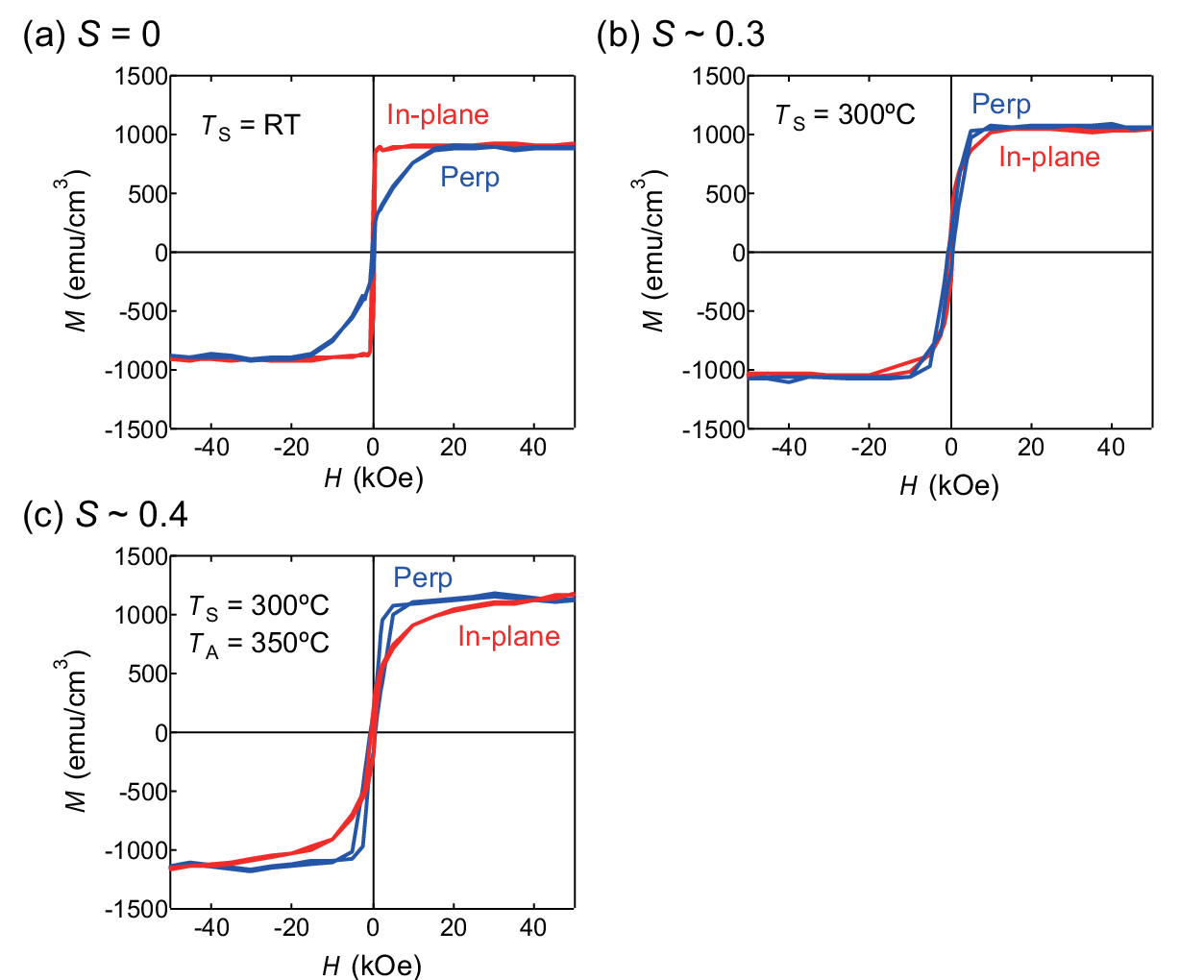}
\caption{Magnetization curves of the grown  $L1_0$ FePt films. The magnetic field is applied parallel or perpendicular to the film plane. 
\label{MHcurves}}
\end{figure}

\end{document}